\documentclass[sigconf,screen,nonacm,review=false,timestamp=false]{acmart}
\AtBeginDocument{%
  }

\usepackage{multirow}
\usepackage{todonotes}
\begin{document}

\title{Learning Factors in AI-Augmented Education: A Comparative Study of Middle and High School Students}

\author{Gaia Ebli}
\authornote{Both authors contributed equally to this research.\\Preprint. Under review.}
\email{gaia.ebli@unibo.it}
\orcid{0009-0000-8382-1338}
\author{Bianca Raimondi}
\authornotemark[1]
\email{bianca.raimondi3@unibo.it}
\orcid{0009-0002-1562-7722}
\author{Maurizio Gabbrielli}
\email{maurizio.gabbrielli@unibo.it}
\orcid{0000-0003-0609-8662}
\affiliation{
  \institution{University of Bologna}
  \city{Bologna}
  \country{Italy}
}

\begin{abstract}
The increasing integration of AI tools in education has led prior research to explore their impact on learning processes. Nevertheless, most existing studies focus on higher education and conventional instructional contexts, leaving open questions about how key learning factors are related in AI-mediated learning environments and how these relationships may vary across different age groups.
Addressing these gaps, our work investigates whether four critical learning factors, experience, clarity, comfort, and motivation, maintain coherent interrelationships in AI-augmented educational settings, and how the structure of these relationships differs between middle and high school students.
The study was conducted in authentic classroom contexts where students interacted with AI tools as part of programming learning activities to collect data on the four learning factors and students' perceptions.
Using a multimethod quantitative analysis, which combined correlation analysis and text mining, we revealed markedly different dimensional structures between the two age groups. Middle school students exhibit strong positive correlations across all dimensions, indicating holistic evaluation patterns whereby positive perceptions in one dimension generalise to others. In contrast, high school students show weak or near-zero correlations between key dimensions, suggesting a more differentiated evaluation process in which dimensions are assessed independently. These findings reveal that perception dimensions actively mediate AI-augmented learning and that the developmental stage moderates their interdependencies.  This work establishes a foundation for the development of AI integration strategies that respond to learners' developmental levels and account for age-specific dimensional structures in student–AI interactions.
\end{abstract}



\maketitle

\section{Introduction}
The integration of Artificial Intelligence (AI) tools in educational contexts has changed the way students learn.
However, the evidence regarding whether dimensions that structure learning experiences operate in the same way when learning is mediated by AI is limited.

Research in educational psychology has established that four dimensions, i.e., experience, clarity, comfort, and motivation, determine learning outcomes~\cite{rodriguez2008students, lo2022students, mokmin2024impact}. Clarity reduces cognitive load and enhances achievement~\cite{serki2024effect, chesebro2001relationship}, comfort supports effective engagement~\cite{aga2024comfort, kiener2014using}, motivation increases initiation and
persistence in activities~\cite{tohidi2012effects}, and positive learning experiences mediate the relationship between pedagogical support and cognitive outcomes~\cite{lo2022students}. These frameworks, however, were developed without AI mediation. With the introduction of conversational, generative, and adaptive AI tools, it is unclear if these dimensional relationships persist and manifest uniformly across developmental stages.

Moreover, while numerous studies have examined the adoption of AI tools, such as ChatGPT and other Large Language Models (LLMs), by university students~\cite{chan2023students, fovsner2024university, grajeda2024assessing}, less attention has been given to younger learners. Studies of K-12 students show awareness of AI benefits alongside concerns about accuracy, overreliance, and ethics~\cite{egara2025secondary, marrone2025understanding, sok2025investigating, higgs2024being}. Critically, most research treats these perceptions as endpoints rather than examining them as interconnected dimensions whose structure may differ developmentally.

This study seeks to answer the following research questions:
\begin{itemize}
    \item RQ1: How do the four dimensions of learning relate to each other in learning contexts mediated by Artificial Intelligence?
    \item RQ2: How do relationships between the four learning dimensions differ between middle and high school students?
\end{itemize} 
Through a multimethod analysis combining quantitative correlation patterns with text mining of student reflections, this study provides empirical evidence of developmental differences in the perception of AI-mediated learning experiences, showing that younger learners exhibit holistic dimensional interplay, where positive perceptions in one dimension generalize to others, while older learners demonstrate differentiation, evaluating each dimension independently based on more sophisticated understanding of AI capabilities.

Our findings have implications for the design of age-appropriate AI educational tools and learning analytics systems, establishing a basis for adaptive interventions that account for age-specific dimensional structures in student–AI interactions, clarifying how AI shapes learning across developmental stages, and guiding the creation of more effective and equitable educational tools.

\section{Related Work}

\subsection{Student Perceptions of Educational Technology}
Research on perceptions of AI in education includes studies concentrating on higher education and diverse disciplines. These works report positive attitudes toward AI effectiveness and usefulness, alongside concerns about accuracy, ethics, privacy, academic integrity, and the need for structured pedagogical integration \cite{grajeda2024assessing, chan2023students, fovsner2024university, arowosegbe2024perception, almufarreh2024determinants}.

Within Computer Science Education (CSE), prior research has specifically investigated generative AI for programming support, code explanation, and problem solving. Studies at the university level show that students perceive AI tools as highly beneficial for learning and productivity, while also raising concerns about overreliance, superficial learning, and occasional inaccuracies \cite{shoufan2023exploring, ma2024enhancing, ahmed2024generative, kohen2025integrating}. These findings align with results from domains such as language learning, writing, and mathematics, suggesting that student perceptions of AI are influenced by general learning processes rather than domain-specific factors alone \cite{wang2024exploring, he2024designing, li2024analyzing, bundgaard2024generative}.

Research on K-12 education remains limited, and most of it doesn't specify if the sample group has any background in CSE. Existing studies indicate that middle and high school students recognize AI’s potential to enhance engagement, comprehension, and problem-solving, while expressing concerns about reliability, ethical implications, privacy, and dependency risks \cite{utami2023utilization, egara2025secondary, marrone2025understanding, sok2025investigating, higgs2024being}. Age-related differences have been observed: chatbot use and confidence increase with age, with high school students reporting greater comfort and perceived productivity benefits than younger learners \cite{klarin2024adolescents}. Existing work on the latter group emphasizes the importance of rules, safety, and scaffolding in AI-supported activities \cite{humburg2024integrating, walan2025primary}.

Overall, prior research has examined single age groups, subjects, or educational levels in isolation, with only a few studies specifically applied to CSE, and almost none of them on K-12 students. Our work adopts a CSE-informed perspective on AI use, comparing middle and high school students to enable a developmental analysis of perceptions of AI-supported learning.

\subsection{Learning Dimensions}
Students’ perceptions of educational technologies are commonly modelled as multidimensional and interrelated. \citet{rodriguez2008students} identified experience, comfort, motivation, and satisfaction as key dimensions of online learning, reporting strong positive correlations between overall experience and individual perception factors. Subsequent work confirmed that learning experiences are central predictors of cognitive outcomes, both directly and indirectly mediated by motivation, self-efficacy, and perceived usefulness \cite{lo2022students, mokmin2024impact}.

Clarity and comfort have been highlighted as particularly influential dimensions. Clarity reduces cognitive load and supports engagement \cite{chesebro2001relationship, serki2024effect}, while comfort, encompassing technological, spatial, and social aspects, facilitates learning \cite{kiener2014using, sharpe2005student, aga2024comfort}.

Despite this body of work, two open questions remain in AI-supported learning: whether traditional correlations persist with AI tools, and whether these relationships differ by developmental stage. This study addresses both gaps by examining interrelationships among learning dimensions and comparing middle and high school students.

\section{Methodology}\label{sec:methodology}

\subsection{Study Design and Context}
The study was conducted in the context of extracurricular schools in Computer Science (CS), where students enrolled in programming courses and encountered AI as the final instructional module.

We adopted a cross-sectional survey design to examine student perceptions of AI tools in authentic CS learning contexts. After informing students about local regulations on personal data protection and requesting and obtaining their informed consent to the processing of their anonymized data, a total of 53 participants were recruited: 25 middle school students (aged 10–13) and 28 high school students (aged 15–20, 11th–12th grade). All participants took part in structured, AI-related activities that were developed and delivered by the authors.

High school students developed Python applications using vibe coding and LLMs such as Gemini \cite{gemini} and Claude \cite{claude}, with an introduction to prompting. Middle school students attended an introductory AI course within a broader CS curriculum, using Teachable Machine \cite{teachablemachine} for hands-on classification and chatbot interaction.

\subsection{Data Collection}
We developed a comprehensive 17-item questionnaire combining quantitative Likert-scale~\cite{likert1932technique} items and open-ended questions to capture measurable perception dimensions.
\begin{table*}[ht]
  \caption{List of closed-ended items with response options}
  \label{tab:questionnaire-closed}
  \begin{tabular}{c p{0.4\linewidth} p{0.48\linewidth}}
    \toprule
    \textbf{ID} & \textbf{Question} & \textbf{Response options} \\
    \midrule
    Q1 & How would you evaluate the overall experience of using the AI tool during the course? & 
    1 = Very poor; 2 = Poor; 3 = Fair; 4 = Good; 5 = Very Good \\
    Q3 & Did you find the AI tool easy to use? & 
    1 = Not at all; 2 = A little; 3 = Quite; 4 = Very much \\
    Q5 & How clear were the AI tool’s responses during the exercises? & 
    1 = Never clear; 2 = Rarely clear; 3 = Occasionally clear; 4 = Frequently clear; 5 = Very frequently clear \\
    Q6 & To what extent did the AI tool help you better understand the course content? & 
    1 = Not at all; 2 = A little; 3 = Quite; 4 = Very much \\
    Q8 & Did you feel comfortable interacting with the AI tool? & 
    1 = Not at all; 2 = Not much; 3 = Partly; 4 = Completely \\
    Q9 & Did you ever have difficulty understanding the AI tool’s responses? & 
    1 = Very frequently; 2 = Frequently; 3 = Occasionally; 4 = Rarely; 5 = Never \\
    Q11 & Would you like to use AI tools in other school courses as well? & 
    1 = I don’t know; 2 = No; 3 = Yes \\
    Q14 & Did using the AI tool motivate you more in doing the exercises? & 
    1 = It demotivated me; 2 = Made no difference; 3 = A little; 4 = Very much \\
    \bottomrule
  \end{tabular}
\end{table*}
Likert-scale questions are reported in Table~\ref{tab:questionnaire-closed}.

Quantitative measures assessed four primary dimensions:
\begin{itemize}
\item Experience: Q1 assessed students' holistic evaluation of their AI tool experience, and Q3 measured ease of use
\item Clarity: Q5 evaluated clarity of AI responses and Q6 assessed learning support
\item Comfort: Q8 measured comfort level and Q9 assessed comprehension difficulties
\item Motivation: Q11 evaluated future usage intentions and Q14 assessed motivational impact
\end{itemize}
\begin{table}[ht]
  \caption{List of open-ended items}
  \label{tab:questionnaire-open}
  \begin{tabular}{c p{0.85\linewidth}}
    \toprule
    \textbf{ID} & \textbf{Question}\\
    \midrule
    Q2 & What were the most useful aspects of using the AI tool during the exercises? \\
    Q4 & Based on how easy or difficult the tool was for you to use, why did you respond this way? \\
    Q7 & In what ways do you think the AI tool influenced your learning? \\
    Q10 & If you had difficulty understanding the AI tool’s responses, can you describe an example or moment? \\
    Q12 & If you would like to use an AI tool in other school courses, why? \\
    Q13 & Do you have suggestions on how to improve the use of AI tools in courses? \\
    Q15 & How did you feel using the AI tool? \\
    Q16 & Is there something you would have liked to do with the AI tool but could not? \\
    Q17 & How do you think AI could be used even more effectively at school? \\
    \bottomrule
  \end{tabular}
\end{table}
The reliability of the instrument was assessed through Cronbach’s alpha \cite{cronbach1951coefficient}, computed across all items as well as separately for middle and high school students. The coefficients were 0.85 (total sample), 0.81 (middle school), and 0.89 (high school), indicating internal consistency and supporting the adequacy of the instrument.

Student opinions were collected using open-ended questions in Table~\ref{tab:questionnaire-open}.
The questionnaire was delivered through Wooclap \cite{wooclap}.

\subsection{Analysis Approach}
Our analysis employed methodologies aligned with best practices:
\begin{itemize}
    \item Statistical Analysis: Correlation matrices, and Mann-Whitney U tests~\cite{nachar2008mann} for group comparisons and descriptive statistics to identify perception patterns and developmental differences.
    \item Text Mining Analysis: Word frequency analysis and co-occurrence network analysis.
\end{itemize}

\section{Results}\label{sec:results}

\subsection{Statistical Analysis}
\begin{table*}[ht]
  \caption{Statistics of closed-ended questions with 5 options.}
  \label{tab:q-5opt}
  \begin{tabular}{llcccccccccc}
    \toprule
    ID & Group & Mean & Median & SD & $U$ & $p$ & Score 1 & Score 2 & Score 3 & Score 4 & Score 5 \\
    \midrule
    \multirow{2}{1.5em}{Q1} & Middle & 3.84 & 4.00 & 0.85 & \multirow{2}{3em}{286.00} & \multirow{2}{2.5em}{0.2106} & 0.00\% & 8.00\% & 20.00\% & 52.00\% & 20.00\% \\
    & High & 4.14 & 4.00 & 0.65 & & & 0.00\% & 0.00\% & 14.30\% & 57.10\% & 28.60\% \\
    \midrule
    \multirow{2}{1.5em}{Q5} & Middle & 3.88 & 4.00 & 0.73 & \multirow{2}{3em}{328.50} & \multirow{2}{2.5em}{0.6859} & 0.00\% & 0.00\% & 32.00\% & 48.00\% & 20.00\% \\
    & High & 3.96 & 4.00 & 0.74 & & & 0.00\% & 0.00\% & 28.60\% & 46.40\% & 25.00\% \\
    \midrule
    \multirow{2}{1.5em}{Q9} & Middle & 3.64 & 4.00 & 1.15 & \multirow{2}{3em}{356.00} & \multirow{2}{2.5em}{0.9172} & 8.00\% & 4.00\% & 28.00\% & 36.00\% & 24.00\% \\
    & High & 3.71 & 4.00 & 0.76 & & & 0.00\% & 3.60\% & 35.70\% & 46.40\% & 14.30\% \\
    \bottomrule
  \end{tabular}
\end{table*}

\begin{table*}[ht]
  \caption{Statistics of closed-ended questions with 4 options.}
  \label{tab:q-4opt}
  \begin{tabular}{llccccccccc}
    \toprule
    ID & Group & Mean & Median & SD & $U$ & $p$ & Score 1 & Score 2 & Score 3 & Score 4 \\
    \midrule
    \multirow{2}{1.5em}{Q3} & Middle & 3.16 & 3.00 & 0.62 & \multirow{2}{3em}{192.00} & \multirow{2}{2.5em}{\textbf{0.0013}} & 4.00\% & 0.00\% & 72.00\% & 24.00\% \\
    & High & 3.68 & 4.00 & 0.48 & & & 0.00\% & 0.00\% & 32.10\% & 67.90\% \\
    \midrule
    \multirow{2}{1.5em}{Q6} & Middle & 2.80 & 3.00 & 0.91 & \multirow{2}{3em}{305.50} & \multirow{2}{2.5em}{0.3773} & 12.00\% & 16.00\% & 52.00\% & 20.00\% \\
    & High & 3.04 & 3.00 & 0.69 & & & 3.60\% & 10.70\% & 64.30\% & 21.40\% \\
    \midrule
    \multirow{2}{1.5em}{Q8} & Middle & 3.40 & 4.00 & 0.96 & \multirow{2}{3em}{297.50} & \multirow{2}{2.5em}{0.2499} & 8.00\% & 8.00\% & 20.00\% & 64.00\% \\
    & High & 3.75 & 4.00 & 0.44 & & & 0.00\% & 0.00\% & 25.00\% & 75.00\% \\
    \midrule
    \multirow{2}{1.5em}{Q14} & Middle & 3.04 & 3.00 & 0.93 & \multirow{2}{3em}{381.50} & \multirow{2}{2.5em}{0.5559} & 8.00\% & 16.00\% & 40.00\% & 36.00\% \\
    & High & 2.96 & 3.00 & 0.74 & & & 0.00\% & 28.60\% & 46.40\% & 25.00\% \\
    \bottomrule
  \end{tabular}
\end{table*}

\begin{table*}[ht]
  \caption{Statistics of closed-ended questions with 3 options.}
  \label{tab:q-3opt}
  \begin{tabular}{llcccccccc}
    \toprule
    ID & Group & Mean & Median & SD & $U$ & $p$ & Score 1 & Score 2 & Score 3 \\
    \midrule
    \multirow{2}{1.5em}{Q11} & Middle & 2.32 & 3.00 & 0.80 & \multirow{2}{3em}{346.00} & \multirow{2}{2.5em}{0.9443} & 20.00\% & 28.00\% & 52.00\% \\
    & High & 2.29 & 3.00 & 0.94 & & & 32.10\% & 7.10\% & 60.70\% \\
    \bottomrule
  \end{tabular}
\end{table*}

\subsubsection{Overall Perception Patterns}
Analysis of closed-ended responses revealed generally positive attitudes toward AI tool usage, with variation across perception dimensions. Tables~\ref{tab:q-5opt}, \ref{tab:q-4opt}, and \ref{tab:q-3opt} present the distribution of responses across all closed-ended questions, stratified by school level.

Middle school students generally displayed slightly more dispersed responses. In Q1 (overall experience), 20.0\% rated their experience as excellent (score 5) and 8.0\% as poor (score 2). High school students showed a more moderate distribution, with 28.6\% selecting the highest rating and 0\% at the lowest scores (1 and 2).

Analysis of the eight quantitative perception items revealed a singular statistically significant difference between the two developmental groups for question Q3 (ease of use) ($U=192.00, \rho=0.0013$). Applying the stringent Bonferroni~\cite{dunn1964multiple} correction ($\alpha_{\text{Bonferroni}} = 0.05/8 = 0.00625$), the p-value for Q3 ($p=0.0013$) remained below this adjusted critical value, confirming its robust statistical significance. This finding underscores a substantial difference in perceived usability between the groups. High school students reported markedly higher ease of use (67.9\% rating the tool as \textit{very much} easy), while middle school students showed considerably more difficulty (only 24.0\% selected \textit{very much} easy). This gap likely reflects older students' greater exposure to conversational AI.

Overall, the results from all questions suggest generally positive experiences across both groups, with middle school students exhibiting slightly more variability, which may reflect less established criteria for evaluating educational technology or variability in baseline digital literacy. The uniform positive distribution among high school students suggests greater consensus and familiarity. These patterns suggest middle school implementations may benefit from more structured introductions, while high school students appear ready for more complex AI interactions.

\begin{figure*}[h]
  \centering
  \includegraphics[width=.85\linewidth]{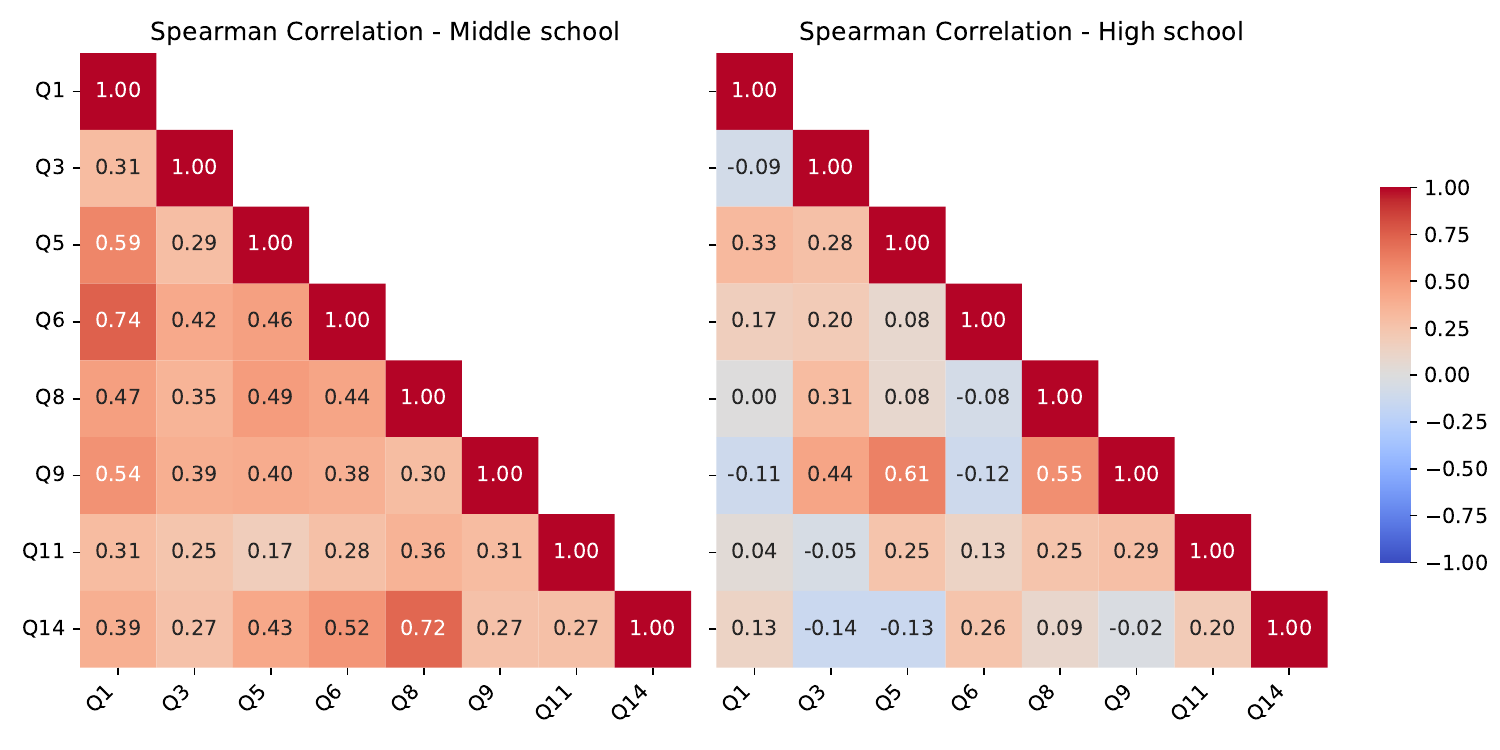}
  \caption{Correlation matrices between questionnaire items by school level, revealing markedly different patterns of association between middle and high school students.}
  \Description{Heatmaps showing correlation patterns between survey items, with middle school students showing predominantly positive correlations while high school students exhibit more complex patterns with several negative correlations.}
  \label{fig:correlation_matrices_by_level}
\end{figure*}

The correlation analysis in Figure~\ref{fig:correlation_matrices_by_level} revealed distinct patterns. Middle school students correlation matrix shows predominantly strong positive correlations (e.g., Q1 and Q6, $\rho = 0.74$; Q8 and Q14, $\rho = 0.72$), indicating that a positive overall perception generalizes to positive evaluations of clarity, usability, and learning support. In contrast, high school students exhibited a more complex pattern with fewer positive correlations. The strongest positive correlation was between Q5 and Q9 ($\rho = 0.61$), both related to clarity. This suggests that older students separate their judgments, and a positive overall experience does not automatically imply a perception of learning impact or motivation.

\subsubsection{Dimension-Level Analysis}
\begin{figure*}[h]
  \centering
  \includegraphics[width=.65\linewidth]{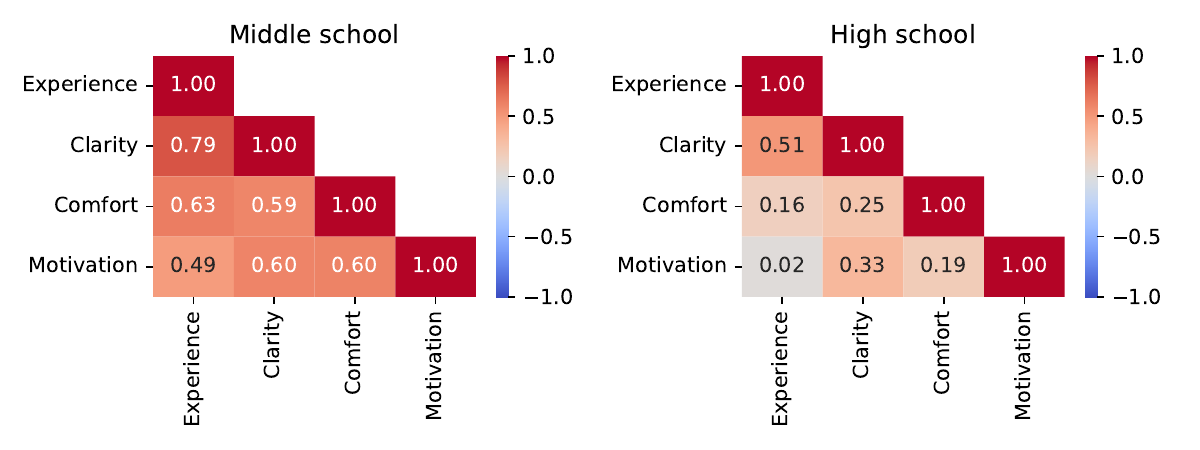}
  \caption{Correlations between aggregated dimensions (Experience, Clarity, Comfort, Motivation) for each school level, showing substantial differences in evaluation coherence between age groups.}
  \Description{Correlation matrix comparing four aggregated dimensions between middle school and high school students, demonstrating that younger students show more holistic evaluation patterns while older students evaluate dimensions more independently.}
  \label{fig:dimension_correlation_matrix}
\end{figure*}
Figure~\ref{fig:dimension_correlation_matrix} displays the correlations between the four aggregated dimensions. For middle school students, all dimension correlations were positive and substantial (e.g., Experience and Clarity, $\rho = 0.79$; Clarity and Motivation, $\rho = 0.60$). This indicates that younger students form a global impression: clear responses likely lead to greater comfort, motivation, and overall satisfaction. The strong coupling of Experience and Clarity suggests that for middle schoolers, understanding the AI’s responses is central to judging the entire interaction.

High school students showed a completely different pattern. While Experience and Clarity maintained a moderate positive correlation ($\rho = 0.51$), several correlations were weak or near zero (Experience and Motivation $\rho = 0.02$; Experience and Comfort $\rho = 0.16$). This decoupling suggests older students separate perceived clarity and usability from motivational engagement. Even if the system is clear and functional, this does not automatically translate into higher motivation or comfort. This independence reflects a more critical and differentiated appraisal style, where each dimension is evaluated on its own merits.

Overall, these patterns highlight a developmental trajectory: middle school students show tightly interconnected perceptions (holistic evaluation), while high school students exhibit a more modular structure (differentiated evaluation). This points to the need for age-sensitive design strategies: emphasizing clarity to boost holistic experience in younger students, while directly addressing motivational factors for older learners.

\subsection{Text Mining Analysis}
\begin{figure*}[h]
  \centering
  \includegraphics[width=0.7\linewidth]{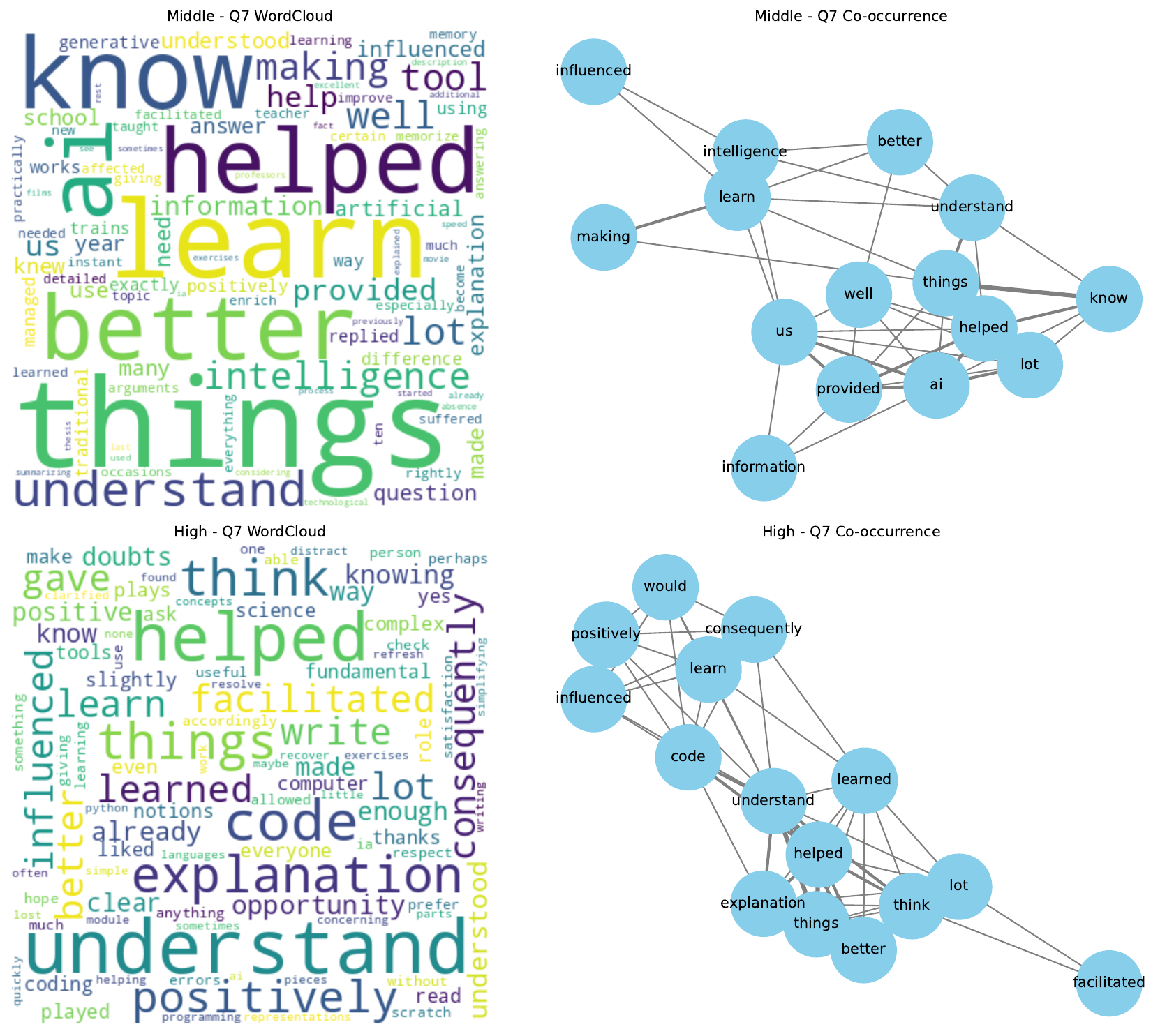}
  \caption{WordCloud and co-occurrence network for Q7 responses, showing student perceptions of the AI tool's impact on learning for middle and high school. The WordCloud highlights the most frequent terms, while the co-occurrence network reveals semantic relationships between key concepts.}
  \Description{Combined visualization with WordClouds and co-occurrence networks for Q7. For middle school, frequent words include "learn," "understand," "better," "helped," "know," with central co-occurrences linking "helped," "know," and "understand." For high school, key words are "understand," "helped," "explanation," "think," "code positively," with "understand" as a central hub in the co-occurrence network connecting the other terms.}
  \label{fig:wordcloud_graph_Q7}
\end{figure*}
Open-ended responses were analyzed separately for middle and high school students to examine their perceptions of the AI tool’s impact on learning. Only the most significant result from this analysis is reported, presented separately for each group.

Figure~\ref{fig:wordcloud_graph_Q7} presents the WordCloud and co-occurrence network for question Q7 (AI influence on learning).
For middle school, the most prominent words in the WordCloud were \textit{learn}, \textit{understand}, \textit{better}, and \textit{helped}, highlighting a focus on comprehension and general learning support. The co-occurrence network reveals a central cluster connecting \textit{helped}, \textit{know}, and \textit{understand}, emphasizing the students’ perception of the AI as a facilitator of understanding and skill acquisition.
For high school, key terms included \textit{understand}, \textit{helped}, \textit{explanation}, \textit{think}, \textit{code}, and \textit{positively}, reflecting a more analytical perspective. In the network, \textit{understand} sits at the center, strongly connected to \textit{helped}, \textit{code}, \textit{explanation}, \textit{think}, and \textit{better}, indicating that students valued the AI both for comprehension and for supporting coding and reasoning tasks.

Overall, this visualization shows the same results of patterns recognized in closed-ended questions: while both groups perceive the AI as beneficial, middle school students emphasize general learning support, whereas high school students focus more on analytical reasoning, explanation, and coding applications.

\section{Discussion}
The four dimensions of perception (Experience, Clarity, Comfort, and Motivation) are significant indicators of learning, consistent with findings from previous studies in traditional educational contexts \cite{chesebro2001relationship, mokmin2024impact, serki2024effect}. Addressing RQ1, our analysis confirms that these dimensions are interconnected through complex relationships that positively influence the students' learning process. Furthermore, the dimensional structure we observed in student-AI interactions shows similar patterns identified in traditional online and classroom learning environments~\cite{rodriguez2008students}. This suggests that the fundamental mechanisms of learning perception remain stable even when mediated by generative AI systems, providing a solid basis for integrating AI tools into educational contexts, as students apply similar learning factors regardless of the support tool's technological nature.

Addressing RQ2, our findings highlight significant developmental differences between middle and high school students that are critical for the design of AI-based educational interventions. The distinct correlation structures indicate that efficient AI-tool integration must necessarily account for developmental stage. Younger students evaluate their experience more holistically: a positive perception in one dimension tends to generalize to others. In contrast, more mature students demonstrate more differentiated and technically sophisticated evaluation patterns, separating usability from motivational impact and clarity from ease of use with greater precision. This distinction suggests that developmental differences should be incorporated into AI-based interventions, adapting implementation strategies to students' level of cognitive maturity.

However, it's important to highlight a couple of potential limitations. First, the context should be considered. Participants were drawn from extracurricular computer science (CS) schools, which are attended by students who are already highly motivated and interested in CS. These students come from the same cultural and educational context, shaped by local curricular practices and regulatory frameworks, which may influence attitudes towards AI and instructional approaches. Second, differences between middle and high school students may reflect developmental factors as well as variations in task complexity, learner autonomy and interaction modalities. The use of different AI tools may also elicit different perceptions and introduce ambiguity in cross-group structural comparisons.

\section{Conclusion and Future Work}
This study provides empirical evidence on incorporating student perception data into educational AI research. By combining statistical correlation analysis with computational text mining, we confirmed the relationship between key learning factors also in AI-mediated learning contexts, but we identified developmental differences in how middle and high school students perceive, evaluate, and interact with AI tools, as questioned in RQs. Middle school students tended to adopt a holistic evaluative perspective, while high school students demonstrated more differentiated and technically sophisticated assessments of AI capabilities and limitations. These findings suggest that student opinion can be used to predict AI adoption and educational effectiveness.

Future work should prioritize several directions. First, differentiated implementation strategies of CSE by developmental level are needed. For younger learners, interventions should privilege clarity and immediate comprehensibility to maximize a positive overall experience. For older students, activities should address usability and technical communication challenges while leveraging their higher-order evaluative capacities. Such differentiation aligns with the goal of personalized and adaptive learning. Second, our findings emphasise the need to explicitly integrate prompt engineering into the CSE AI module curriculum to enhance students' ability to benefit fully from AI systems. This is because during lessons, we found difficulties in students formulating effective prompts. Third, the effective adoption of these innovations depends on teacher professional development. Training programs should prioritize both AI literacy and practical competencies in prompt engineering. Finally, we propose a gradual introduction of AI tools that respects cognitive developmental stages, using structured templates for younger students and promoting autonomy for older learners. In addition, longitudinal studies will be crucial to capture how student perceptions evolve as AI literacy develops. Integrating perception data with behavioral usage traces will provide a more comprehensive picture of student–AI interaction dynamics. By pursuing these directions, future work can contribute to the design of AI-supported learning environments that are both pedagogically effective and developmentally responsive.

\bibliographystyle{ACM-Reference-Format}
\bibliography{sample-base}

\end{document}